\documentclass[a4paper,scale=0.8,onecolumn,nobibnotes,nofootinbib]{revtex4}
\usepackage{enumitem}
\usepackage{amsmath,amssymb}
\usepackage{graphicx} 
\usepackage{epstopdf}
\usepackage{hyperref}
\usepackage{geometry}
\usepackage{subfigure}

\hypersetup{
	colorlinks=true,
	citecolor=blue,
	linkcolor=red,
	filecolor=magenta,
	urlcolor=cyan,}

\newcommand{\be}{\begin{equation}}
\newcommand{\ee}{\end{equation}}

\begin{document}
	
	\title{\large\bf Thermodynamical properties of a deformed Schwarzschild black hole via Dunkl generalization}
	\author{P. Sedaghatnia
		\footnote{E-mail: pa.sedaghatnia@gmail.com}$^{,1}$,\,
		H. Hassanabadi
		\footnote{E-mail: h.hasanabadi@shahroodut.ac.ir }$^{,2}$,\, A. A. Araújo Filho
		\footnote{E-mail: dilto@fisica.ufc.br (The corresponding author) }$^{,3,4}$,\,P. J. Porf\'{\i}rio\footnote{E-mail: pporfirio@fisica.ufpb.com}$^{,4}$
		and
		W. S. Chung
		\footnote{E-mail: mimip44@naver.com}$^{,5}$
	}
	\affiliation{$^1$ Faculty of Physics, Shahrood University of Technology, Shahrood, Iran P. O. Box : 3619995161-316.\\
		$^2$ Department   of   Physics, Faculty of Science,  University   of   Hradec   Kr\'{a}lov\'{e}, Rokitansk\'{e}ho   62, 500   03   Hradec   Kr\'{a}lov\'{e},   Czechia \\
		$^3$ Departamento de Física Teórica and IFIC, Centro Mixto Universidad de Valencia--CSIC. Universidad de Valencia, Burjassot-46100, Valencia, Spain. \\
		$^4$ Departamento de Física, Universidade Federal da Paraíba, Caixa Postal 5008, 58051-970, João Pessoa, Paraíba,  Brazil. \\
        $^5$ Department of Physics and Research Institute of Natural Science, College of Natural Science, \\Gyeongsang National University, Jinju 660-701, Korea}
	\date{\today}

\begin{abstract}
In this paper, we construct a deformed Schwarzschild black hole from the de Sitter gauge theory of gravity within Dunkl generalization and we determine the metric coefficients versus Dunkl parameter and parity operators. 
Since the spacetime coordinates are not affected by the group transformations, only fields are allowed to change under the action of the symmetry group. A particular ansatz for the gauge fields is chosen and the components of the strength tensor are computed as well. Additionally, we analyze the modifications on the thermodynamic properties to a spherically symmetric black hole due to Dunkl parameters for even and odd parities. Finally, we verify a novel remark highlighted from heat capacity: the appearance of a phase transition when the odd parity is taken into account.
\newline\newline	  
{\it Keywords}: Dunkl Operator; Gauge Theory; Thermodynamic Properties.
\end{abstract}
\maketitle
	
\section{Introduction}\label{sec1}
	
	The Dunkl operator in spherical coordinates has received much attention over the last years \cite{1,2,3,4,5,6,7,8,9,10,11,12,13,14}. Essentially, such an operator consists of three parts. One of them has the normal derivative term, while the other two ones include parity, which can be written into two different manners: even and odd parity states. In fact, it represents a meaningful generalization of partial derivatives, since there is an involvement of differential operators and also finite reflection groups in order to provide a concise structure for multi--variable analysis. Under a certain limit case for the Dunkl deformation parameters, the initial form of the normal derivatives is naturally recovered.

	More so, recently, many works have been performed with the purpose of developing a gauge theory of gravitation \cite{15,16}. This approach is a theory of general relativity, in the de Sitter group SO(4,1), on a commutative 4--dimensional metric with a spherical symmetry in Minkowski spacetime \cite{1}:
	\begin{eqnarray}\label{1}
	\mathrm{d}s^{2}=\mathrm{d}t^{2}-\mathrm{d}r^{2}-r^{2}\left(\mathrm{d}\theta^{2}+\sin^{2}\theta \,\mathrm{d}\phi^{2}\right).
	\end{eqnarray} 
	The de Sitter group is a 10--dimensional one, where the gravitational field is described by gauge field potentials $h_{\mu}^{A}$, $\mu=0,1,2,3$ and $A=1,2,..,10$. These ones depend on the coordinates of the base manifold and they are split into six spin connections $\omega_{\mu}^{ab}(x)$ and four tetrad fields $e_{\mu}^{a}(x)$. Such a group is identified with $\omega_{\mu}^{ab}(x)=-\omega_{\mu}^{ba}(x)$ and $\omega_{\mu}^{a5}(x)=\lambda e_{\mu}^{a}(x)$, where $\lambda$ is a contraction parameter. According to Refs. \cite{1,15,16,17,18}, using the tetrad and the spin connections, the strength tensor components is introduced as follows:
	\begin{eqnarray}\label{2}
	&&
	F_{\mu \nu}^a= \partial_\mu e_\nu^a-\partial_\nu e_\mu^a+\left(\omega_\mu^{a b} e_\nu^c-\omega_\nu^{a b} e_\mu^c\right) \eta_{b c}=T_{\mu \nu}^a \\
	&&
	F_{\mu \nu}^{a b}= \partial_\mu \omega_\nu^{a b}-\partial_\nu \omega_\mu^{a b}+\left(\omega_\mu^{a c} \omega_\nu^{d b}-\omega_\nu^{a c} \omega_\mu^{d b}\right) \eta_{c d}-4 \lambda^2\left(e_\mu^a e_\nu^b-e_\nu^a e_\mu^b\right)=R_{\mu \nu}^{a b}
	\end{eqnarray}
	where $\eta_{ab}=diag\left(1,-1,-1,-1\right)$.
	On the other hand, the field equations for the gravitational gauge potentials $\omega_{\mu}^{a}(x)$ are $F_{\mu \nu}^a=0$, where there exists the absence of torsion.

	In this paper, we construct a deformed Schwarzschild black hole from the de Sitter gauge theory of gravity in the presence of Dunkl formalism. After that, we determine the metric coefficients with respect to Dunkl parameter and parity operators.
	Next, to corroborate our results, we apply them to investigate the respective thermodynamic properties of the system.

	This paper is organized as follows: in Sec. \ref{sec2}, we review the formalism of Dunkl operator on spherical coordinates, and we use this approach to solve explicitly the field equations for a specific case of the deformed Schwarzschild spacetime. Also, we recover the well--known deformed Schwarzschild solution with a cosmological constant when Dunkl parameters vanish. In Sec. \ref{sec3} a particular ansatz for the gauge fields is chosen and the components of the strength tensor are computed. Also, this formalism allows us to find the Riemann tensor, the Ricci tensor, the curvature scalar, the field equations, and the integration of these equations according to Dunkl parameters and parity operators. In Sec. \ref{sec4}, we analyze the modifications on the thermodynamical properties of the black hole due to the Dunkl contributions for even and odd parities, and we show that they play an important role in removing critical points. In the last Sec. \ref{Conclusions},  we present our remarks and conclusions.
	
	
	\section{Dunkl operator in Spherical coordinates}\label{sec2}

	Let us introduce the general form of the Dunkl operator first in Cartesian coordinates as follows \cite{2,3,4,5,6,7,8,9,10}:
	\begin{eqnarray}\label{4}
	D_{x_{i}}= \dfrac{\partial}{\partial x_{i}}+\dfrac{\alpha_{i}}{x_{i}}(1-\mathcal{R}_{i})\quad,\quad(i=0,1,2,3)
	\end{eqnarray}
	where $\alpha_{i}=\left(0,\alpha_{1},\alpha_{2},\alpha_{3}\right)$  and $\mathcal{R}_{i}=\left(0,\mathcal{R}_{1},\mathcal{R}_{2},\mathcal{R}_{3}\right)$ are Dunkl parameters $(\alpha_{i}> -1/2)$, and the parity operators, respectively. The Dunkl operator formalism incorporates parity transformations and finite reflection symmetries into the study of spacetime geometries. These deformations extend beyond mathematical generalizations, with potential to reveal new features, especially in contexts where parity violation or reflection symmetry breaking plays a significant role. Such scenarios include quantum gravity frameworks and theories addressing chiral anomalies \cite{11,12,13,14}. Recently, the Dunkl derivative was introduced in flat space, by redefining spatial derivatives with respect to $x$, $y$, and $z$, through the introduction of reflection operators. Another important remark relies on the spherical coordinates: these ones produce a modification of planar angle on the reflection operator. In this sense, the parity operators are written as follows \cite{11}: 
	\begin{eqnarray}\label{5}
	\mathcal{R}_{1} f(t,r,\theta,\phi)=f(t,r,\theta,\pi-\phi),\quad
	\mathcal{R}_{2} f(t,r,\theta,\phi)=f(t,r,\pi-\theta,\phi),\quad
	\mathcal{R}_{3} f(t,r,\theta,\phi)=f(t,r,\theta,-\phi).
	\end{eqnarray}
	Changing the coordinates from Cartesian to spherical ones and observing the properties of Eq. \eqref{5}, the components of Dunkl operators in spherical coordinates read \cite{40}
	\begin{equation}\label{6}
	\begin{aligned}
	&D_{r}=\frac{\partial}{\partial r}+\frac{1}{r}\sum_{i=1}^{3}\alpha_{i}(1-\mathcal{R}_{i}),\quad\quad D_{t}=\frac{\partial}{\partial t},\\
	&D_{\theta}=\frac{\partial}{\partial \theta}+\sum_{i=1}^{2}\alpha_{i}(1-\mathcal{R}_{i}) \cot\theta-\alpha_{3}(1+\mathcal{R}_{3})\tan\theta,\\
	&D_{ \phi}=\frac{\partial}{\partial \phi}-\alpha_{1} \tan\phi (1-\mathcal{R}_{1})+\alpha_{2} \cot\phi (1-\mathcal{R}_{2}).
	\end{aligned}
	\end{equation}
	In the next section, based on theses features highlighted so far, we shall provide an investigation on the deformed Schwarzschild spacetime from the de Sitter gauge theory of gravity in the presence of Dunkl operators.
	
	\section{The deformed Schwarzschild spacetime in the presence of Dunkl operators}\label{sec3}

	In order to accomplish our study of such a modified spacetime, we consider a particular form of the diagonal tetrad fields by using the ansatz \cite{ 15,16,17,18}:
	\begin{align}\label{7}
	e_\mu^0=(A, 0,0,0), \quad e_\mu^1=\left(0, \frac{1}{A}, 0,0\right), \quad
	e_\mu^2=(0,0, r\, C, 0), \quad e_\mu^3=(0,0,0, r\, C \sin \theta), 
	\end{align}
	together with the following spin connections \cite{15}
	\begin{align}\label{8}
	\omega_\mu^{01}=(U, 0,0,0), \quad \omega_\mu^{12}=(0,0, W, 0), \quad \omega_\mu^{13}=(0,0,0, Z \sin \theta), \quad
	\omega_\mu^{23}=(V, 0,0, \cos \theta),
	\end{align}
	where $A, C, U, W, Z$ and $V$ are functions of the radial coordinate $r$. The Einstein equations for the vacuum case read \cite{15,16}
	\begin{eqnarray}\label{9}
	\tilde{R}_{\mu}^{\nu}-\frac{1}{2}\delta_{\mu}^{\nu}\tilde{R}=0,
	\end{eqnarray}
	where $\tilde{R}_{\mu}^{\nu}$ and $\tilde{R}$ are the Ricci tensor and the Ricci scalar for our model (see Appendix \ref{ApendixB}). More so, their respective field equations may also be calculated as follows:
	\begin{eqnarray}\label{10}
	\frac{\left(1-W Z+\zeta\right)}{r^2 C^2}+
	\frac{(W+Z) A}{r^2 C} \sum_{i=1}^3 \alpha_i\left(1-\mathcal{R}_{i}\right)-\frac{\left(W^{\prime}+Z^{\prime}\right) A}{r C}+12 \lambda^2=0,
	\end{eqnarray}
	\begin{eqnarray}\label{11}
	U^{\prime}+\frac{1}{r} \sum_{i=1}^3 \alpha_i\left(1-\mathcal{R}_{i}\right) U+\frac{Z U}{r C A}-\frac{Z^{\prime} A}{r C}-\frac{Z A}{r^2 C} \sum_{i=1}^3 \alpha_i\left(1-\mathcal{R}_{i}\right)+12 \lambda^2=0,
	\end{eqnarray}
	\begin{eqnarray}\label{12}
	U^{\prime}+\frac{1}{r} \sum_{i=1}^3 \alpha_i\left(1-\mathcal{R}_{i}\right) U+\frac{W U}{r C A}-\frac{W^{\prime} A}{r C}-\frac{W A}{r^2 C} \sum_{i=1}^3 \alpha_i\left(1-\mathcal{R}_{i}\right)+12 \lambda^2=0,
	\end{eqnarray}
	\begin{eqnarray}\label{13}
	(W-Z)A+\sum_{i=1}^3 \alpha_i\left(1-\mathcal{R}_{i}\right) \frac{A \cos \theta}{r C \sin \theta}+\alpha_3\left(1+R_3\right) \frac{A \tan \theta}{r C}=0,
	\end{eqnarray}
	\begin{eqnarray}\label{14}
	\frac{(W+Z) U}{r C A}+\frac{1-W Z+\zeta}{r^2 C^2}+12 \lambda^2=0,
	\end{eqnarray}
	where $\zeta=-\alpha_3\left(1+\mathcal{R}_{3}\right)+\sum_{i=1}^2 \alpha_i\left(1-\mathcal{R}_{i}\right) \cot^2\theta$, and the prime symbol ($\prime$) represents the derivative with respect to the radial component $r$. Using those expressions presented in Appendix \ref{A1} with the condition of null--torsion $F_{\mu \nu}^{a}=0$, we obtain the respective constraints \cite{1,15,16}:
	\begin{align}\label{15}
	U=-A A^{\prime}-\frac{A}{r} \sum_{i=1}^3 \alpha_i\left(1-\mathcal{R}_{i}\right),\quad W=Z=A\left(1+\sum_{i=1}^3 \alpha_i\left(1-\mathcal{R}_{i}\right)\right),\quad V=0,\quad C=1.
	\end{align}
	Notice that the combination of Eq. \eqref{11} with Eq. \eqref{12} turns out to cast the same field equation if comprared with Eq. \eqref{10}, and Eq. \eqref{14} by considering the constraints displayed in Eq. (\ref{15}). In this way, we obtain only two independent equations:
	\begin{eqnarray}\label{16}
	\frac{1}{2}\left(A^2\right)^{\prime \prime}+\alpha_{3}(1+\mathcal{R}_{3})-\sum_{i=1}^{2}\alpha_{i}(1-\mathcal{R}_{i})\cot\theta^{2}=0,
	\end{eqnarray}
	\begin{eqnarray}\label{17}
	-\frac{1}{r^{2}}+\frac{A^2}{r^{2}}\left(1+\sum_{i=1}^3 M_i\left(1-\mathcal{R}_{i}\right)\left(1+M_i\left(1-\mathcal{R}_{i}\right)\right)\right)+\alpha_{3}(1+\mathcal{R}_{3})-\sum_{i=1}^{2}\alpha_{i}(1-\mathcal{R}_{i})\cot\theta^{2}=0.
	\end{eqnarray}
	If we take the difference between Eqs. \eqref{16} and \eqref{17},
	we obtain one single expression
	\begin{eqnarray}\label{18}
	r^{2}\left(A^{2}\right)''-2 A^{2}\left(1+\sum_{i=1}^3 \alpha_i\left(1-\mathcal{R}_{i}\right)\left(1+\alpha_i\left(1-\mathcal{R}_{i}\right)\right)\right)+2=0.
	\end{eqnarray}
	Above expression has the following solution:
	\begin{align}\label{19}
	A^2=\frac{1}{\left(1+\xi\right)}+\beta\,r^{\frac{1}{2}\left(1+\sqrt{9+8 \xi}\right)}+\gamma\, r^{\frac{1}{2}\left(1-\sqrt{9+8 \xi}\right)},
	\end{align}
	where $\xi=\sum_{i=1}^3 \alpha_i\left(1-\mathcal{R}_{i}\right)\left(1+\alpha_i\left(1-\mathcal{R}_{i}\right)\right)$ and $\gamma, \beta$ are two arbitrary constants. To accomplish our calculations, we have set $\gamma=-2M$, $\beta=4 \lambda^2$ and $\Lambda=-12 \lambda^2$, where $\Lambda$ is the cosmological constant, and $M$ is the mass of the point--like source of the gravitational field. When the limit $\alpha_{i}=0$ occurs, we recover the black hole solution of the model studied in Ref. \cite{1}
	\begin{align}\label{21}
	A^2=1-\frac{2 M}{r}-\frac{\Lambda}{3 }\,r^{2},
	\end{align}
	where it represents the Schwarzschild de Sitter black hole. The corresponding metric $g_{\mu \nu}=\eta_{a b} e_\mu^a e_\nu^b $ has the following non--zero components 
	\begin{eqnarray}\label{22}
	g_{00}=\frac{-1}{g_{11}}=f(r)=\frac{1}{\left(1+\xi\right)}-2 M\, r^{\frac{1}{2}\left(1-\sqrt{9+8 \xi}\right)}-\frac{\Lambda}{3}\,r^{\frac{1}{2}\left(1+\sqrt{9+8 \xi}\right)} \quad, \quad g_{22}=\frac{g_{33}}{\sin^{2} \theta }=-r^{2}.
	\end{eqnarray}
	
\begin{figure}[h!]
		\centering
		\subfigure[]{
			\begin{minipage}[t]{0.5\linewidth}
				\centering
				\includegraphics[width=3.4in,height=2.6in]{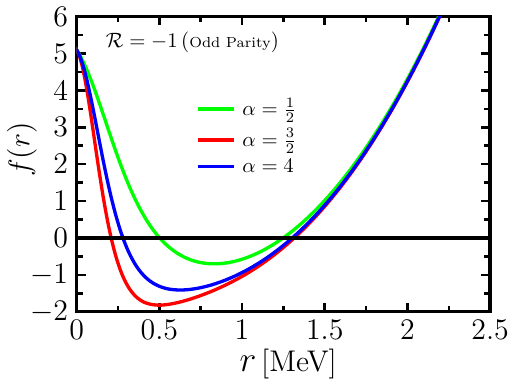}
			\end{minipage}%
		}%
		\subfigure[]{
			\begin{minipage}[t]{0.5\linewidth}
				\centering
				\includegraphics[width=3.4in,height=2.6in]{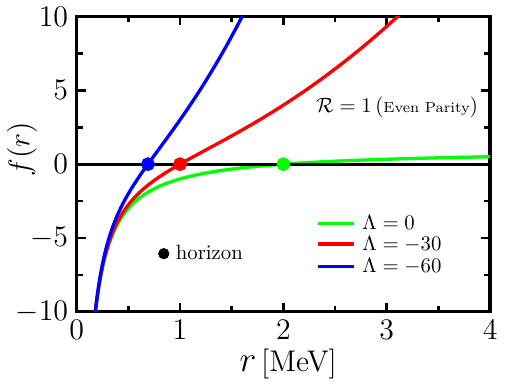}
			\end{minipage}%
		}%
		\centering
		\caption{Function $f(r)$ with respect to $r$ for  (a)  $\mathcal{R}=-1$ (odd parity) with different values of $\alpha=1/2,3/2,4$, and (b) $\mathcal{R}=+1$ (even parity) with different values of $\Lambda=0,-30,-60$. Here, we set $M = 1$.}
		\label{fig1}
	\end{figure}

		In order to obtain the solution to this case, we assume the following form for the line element,
		\begin{eqnarray}\label{22d}
		\mathrm{d}s^{2}=f(r) \mathrm{d}t^{2}-\frac{1}{f(r)}\mathrm{d}r^{2}-r^{2}\left(\mathrm{d}\theta^{2}+\sin^{2} \theta \mathrm{d} \phi^{2}\right).
		\end{eqnarray}
		It is straightforward to see that the metric function Eq. \eqref{22d} turns back to the Schwarzschild black hole solution for the case where $\alpha_{i}=\Lambda=0$.
	As it is well--known, in possession to $f(r)$, we can ensure about the event horizon of a given black hole. In Fig. \eqref{fig1}, we plot the corresponding behavior of $f(r)$ with respect to $r$ for different values of $\alpha$, and parity operators $\mathcal{R}=+1$ (even) and $\mathcal{R}=-1$ (odd). It should be noted that when Dunkl parameter is $\alpha = 0$, we have the well--known case of de Sitter Schwarzschild black hole. Also, from Fig. \eqref{fig1}, we can see that the event horizon radius $r_{H}$ increases whenever there is an increment of Dunkl parameters $\alpha_{i}$ for both parities $\mathcal{R}=\pm 1$. In the literature, a similar situation occurred as well to the black hole proposed by Moffat \cite{41}. Furthermore, it is important to mentioning that the de Sitter Schwarzschild black hole for certain configurations presents one more horizon in comparison with the Schwarzschild black hole before reaching the extreme case \cite{42}.
	
	\section{Thermodynamics}\label{sec4}
	
		In this section, we use the results of the previous sections to study the thermodynamics of the Dunkl--deformed black hole from de Sitter gauge gravity. The first step to accomplish this relies on the calculation of the Dunkl--deformed black hole mass. 
		By solving $f(r)|_{r=r_{H}}=0$ in Eq. \eqref{22}, we can get the relation between the black hole mass $\mathrm{M}_{Dunkl}$ and the event horizon radius $r_{H}$ as follows
	
	\begin{eqnarray}\label{23}
	\mathrm{M}_{Dunkl} = -\frac{r_{H}^{\frac{1}{2} \left(\sqrt{8 \xi+9}-1\right)} \left(\Lambda  (\xi+1) r_{H}^{\frac{1}{2} \left(\sqrt{8 \xi+9}+1\right)}-3\right)}{6 (\xi+1)},
	\end{eqnarray}
	where $H=(\pm)$. Assuming the particular limits where $\alpha_{i}=0$, and $\Lambda=0$, we obtain respectively
	\begin{eqnarray}
	\lim\limits_{\alpha_{i}\rightarrow 0}\mathrm{M}_{Dunkl}=\frac{r_{H}}{2}-\frac{\Lambda\,  r_{H}^{3}}{6}, \,\,\,\,\,\,\,\,\,\, \lim\limits_{\alpha_{i},\Lambda\rightarrow 0}\mathrm{M}_{Dunkl}=\frac{r_{H}}{2}.
	\end{eqnarray}
	In addition, the Hawking temperature is a straightforward task to be accomplished from the surface gravity at the horizon
	\begin{eqnarray}\label{26}
	T=\frac{f'(r_{H})}{4 \pi}
	\end{eqnarray}
	and using above relations, we acquire
	\begin{eqnarray}\label{27}
	T = \frac{-M \left(1-\sqrt{8 \xi +9}\right) r_{H}^{\frac{1}{2} \left(1-\sqrt{8 \xi +9}\right)-1}-\frac{1}{6} \Lambda  \left(\sqrt{8 \xi +9}+1\right) r_{H}^{\frac{1}{2} \left(\sqrt{8 \xi +9}+1\right)-1}}{4 \pi }.
	\end{eqnarray}
	Notice that, in the absence of Dunkl operator $\alpha_{i}=0$, and cosmological constant, we have
	\begin{eqnarray}
	\lim\limits_{\alpha_{i},\, \Lambda\rightarrow 0}T=\frac{\mathrm{M}}{2 \pi  r_H^2}=\frac{1}{4 \pi  r_H}.
	\end{eqnarray}
	
\begin{figure}[h!]
		\centering
		\subfigure[]{
			\begin{minipage}[t]{0.5\linewidth}
				\centering
				\includegraphics[width=3.4in,height=2.6in]{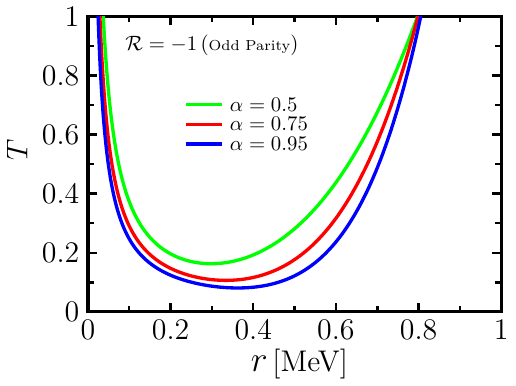}
			\end{minipage}%
		}%
		\subfigure[]{
			\begin{minipage}[t]{0.5\linewidth}
				\centering
				\includegraphics[width=3.4in,height=2.6in]{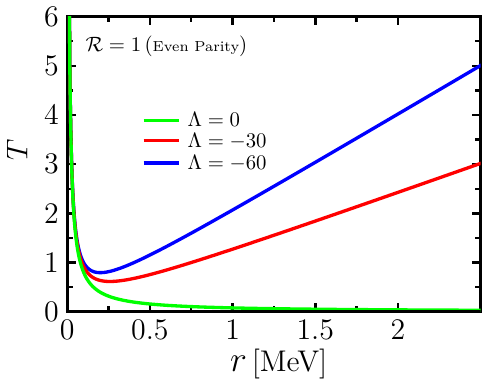}
			\end{minipage}%
		}%
		\centering
		\caption{Plot of temperature $T$ as the function of horizon radius $r$ for (a)  $\mathcal{R}=-1$ (odd parity) with different values of $\alpha=0.5,0.75,0.95$, and (b) $\mathcal{R}=+1$ (even parity) with different values of $\Lambda=0,-30,-60$. }
		\label{fig2}
	\end{figure}

\begin{figure}[h!]
		\centering
		\subfigure[]{
			\begin{minipage}[t]{0.5\linewidth}
				\centering
				\includegraphics[width=3.3in,height=2.6in]{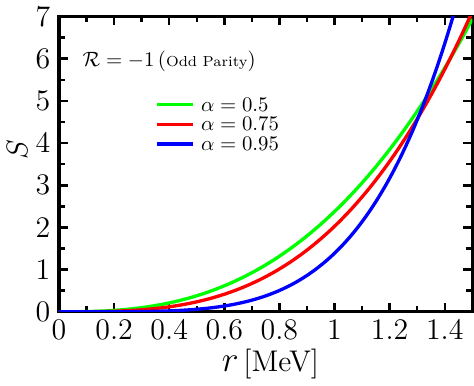}
			\end{minipage}%
		}%
		\subfigure[]{
			\begin{minipage}[t]{0.5\linewidth}
				\centering
				\includegraphics[width=3.3in,height=2.6in]{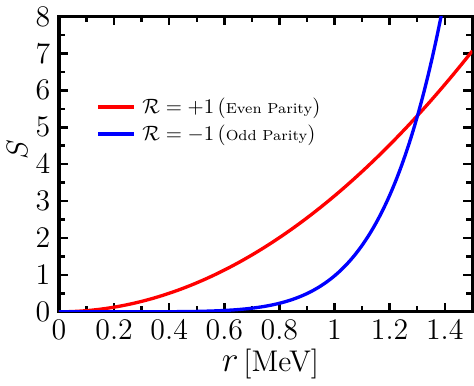}
			\end{minipage}%
		}%
		\centering
		\caption{Entropy $S$ as a function of radius $r$ for (a)  $\mathcal{R}=-1$ (odd parity) with different values of $\alpha=0.5,0.75,0.95$, and (b) $\mathcal{R}=\pm1$ (even and  parity) with $\alpha=0.95$.}
		\label{fig3}
	\end{figure}

		In Fig. \eqref{fig2}, we show the variation of the temperature to (a) $\mathcal{R}=-1$ (odd parity) with different values of Dunkl parameter  $\alpha=1/2,3/2,4$, and (b) $\mathcal{R}=+1$ (even parity) with different values of $\Lambda=0,-30,-60$. Here, we also realize that the temperature is a decreasing function when the radius $r$ increases until attaining its minimum value for a fixed value of Dunkl parameter. It is worthy to be highlighted that the minimum values will occur at $r = r_{H}$, i.e., where the heat capacity diverges. 
	On the other hand, the entropy of Dunkl--deformed black hole can be derived as
	\begin{eqnarray}
	\mathrm{d} S=\frac{1}{T}\bigg(\frac{\partial \mathrm{M}_{Dunkl}}{\partial r_{H}}\bigg)\mathrm{d} r_{H}.
	\end{eqnarray}
	From Eqs. \eqref{23} and \eqref{27}, we have 
	\begin{eqnarray}\label{31}
	\frac{\partial \mathrm{M}_{Dunkl}}{\partial r_{H}}=\frac{r_{H}^{\frac{1}{2} \left(\sqrt{8 \xi +9}-3\right)} \left(3 \left(\sqrt{8 \xi +9}-1\right)-2 \Lambda  (\xi +1) \sqrt{8 \xi +9} r_{H}^{\frac{1}{2} \left(\sqrt{8 \xi +9}+1\right)}\right)}{12 (\xi +1)}
	\end{eqnarray}
	and 
	\begin{eqnarray}\label{32}
	\mathrm{d}S=\left(2 \pi  r_{H}^{\frac{1}{2} \left(\sqrt{8 \xi +9}-1\right)}\right)\mathrm{d}r_{H}.
	\end{eqnarray}
	Then, by integrating these results into Eq. \eqref{32}, we get
	\begin{eqnarray}\label{33}
	S=\frac{4 \pi  r_{H}^{\frac{1}{2} \left(\sqrt{8 \xi +9}+1\right)}}{\sqrt{8 \xi +9}+1}+S_{0},
	\end{eqnarray}
	where $S_{0}$ is the integration constant. In the absence of Dunkl operators, the relation obtained to the entropy reduces to 
	\begin{eqnarray}
	\lim\limits_{\alpha_{i}\rightarrow 0}S_{\alpha_{i}}=\pi \,r_{H}^{2}+S_{0}.
	\end{eqnarray}
Note that above expression accounts for the  entropy for the usual de Sitter Schwarzschild black hole. More so, in Fig. \eqref{fig3}, we depict the entropy $S(r)$ for (a) $\mathcal{R}=-1$ (odd parity) with different values of $\alpha=0.5,0.75,0.95$, and (b) $\mathcal{R}=\pm1$ (even and odd parity) with  $\alpha=0.95$. As it is exhibited in (b), the entropy diverges at $r_{H}\rightarrow 1.4$ for $\mathcal{R}=-1$. We may see that this thermodynamic function is a monotonically increasing function as $r_{H}$ increases. Only at the regions where there exist positive temperatures and positive slopes, the black hole can thermodynamically be stable. In other words, this means that when the black hole and its corresponding cosmological horizon are either sufficiently far (small $r$) or sufficiently close (large $r$) to each other, the stability no longer belongs to our system under consideration.
	In addition, we have
	\begin{eqnarray}
	r_{H}=(4 \pi )^{-\frac{2}{\sqrt{8 \xi +9}+1}} \left(\left(\sqrt{8 \xi +9}+1\right) (\mathrm{S}-\mathrm{S}_{0})\right)^{\frac{2}{\sqrt{8 \xi +9}+1}},
	\end{eqnarray}
	where in the absence of the Dunkl operator, we reach the ordinary radius form as
	\begin{eqnarray}
	\lim\limits_{\alpha_{i}\rightarrow 0}r_{H}=\sqrt{\frac{(\mathrm{S}-\mathrm{S}_{0})}{\pi }}.
	\end{eqnarray}
	According to $F=\mathrm{M}-TS$, we can obtain the Helmholtz free energy
	\begin{eqnarray}
	F=\frac{r_{H}^{\frac{1}{2} \left(\sqrt{8 \xi +9}-1\right)} \left(3-\Lambda  (\xi +1) r_{H}^{\frac{1}{2} \left(\sqrt{8 \xi +9}+1\right)}\right)}{6 (\xi +1)}-T \left(\frac{4 \pi  r_{H}^{\frac{1}{2} \left(\sqrt{8 \xi +9}+1\right)}}{\sqrt{8 \xi +9}+1}+S_{0}\right).
	\end{eqnarray}
	Here, notice that if we assume $\alpha_{i}=0$ and $\Lambda=0$, above expression reads
	\begin{eqnarray}
	\lim\limits_{\alpha_{i},\, \Lambda\rightarrow 0}F=\frac{r_{H}}{2}- T \left(\pi  r_{H}^2\right).
	\end{eqnarray}
	
\begin{figure}[h!]
		\centering
		\subfigure[]{
			\begin{minipage}[t]{0.5\linewidth}
				\centering
				\includegraphics[width=3.4in,height=2.6in]{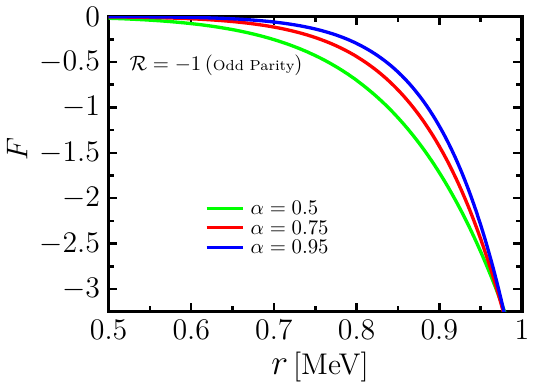}
			\end{minipage}%
		}%
		\subfigure[]{
			\begin{minipage}[t]{0.5\linewidth}
				\centering
				\includegraphics[width=3.4in,height=2.6in]{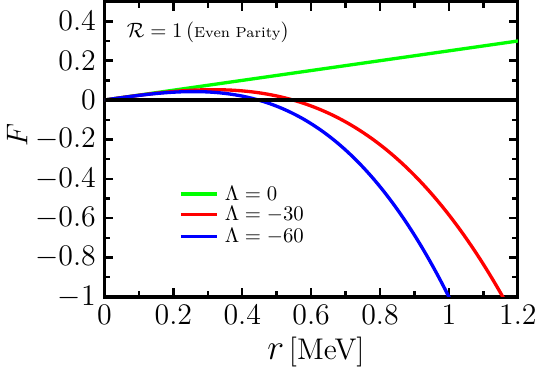}
			\end{minipage}%
		}%
		\centering
		\caption{ Helmholtz free energy $F$ as the function of radius $r$ for (a)  $\mathcal{R}=-1$ (odd parity) with different values of $\alpha=0.5,0.75,0.95$, and (b) $\mathcal{R}=+1$ (even parity) with $\Lambda=0,-30,-60$.}
		\label{fig4}
	\end{figure}

		From Fig. \eqref{fig4}, it is shown that the radius $r$ plays an important role in the physical meaning of free energy. Until $r_{H} = 0.01$ for $\mathcal{R}=-1$, there is an absorption of energy (positive energy). However, after this point such thermodynamic function starts to decrease monotonically. On the other hand, for $\mathcal{R} = +1$, the Helmholtz free energy increases until $r_{H} = 0.2$; also, it is important to realize the role of parameter $\Lambda$ to this thermal quantity. Taking into account a general panorama, we observe that by increasing the radius $r$, the black hole will have negative values of the Helmholtz free energy.
	
 Here, another quantity worthy to be calculated is the pressure. In this sense, we use the relation showed below
	\begin{eqnarray}
	P=-\frac{1}{4 \pi r_{H}^{2}}\bigg(\frac{\partial F}{\partial r_{H}}\bigg)_{T}.
	\end{eqnarray}
	With it, the full equation of states can be derived:
	\begin{eqnarray}
	P=\frac{r_{H}^{\frac{1}{2} \left(\sqrt{8 \xi +9}-7\right)} \left(-3 \sqrt{8 \xi +9}+2 \Lambda  (\xi +1) \sqrt{8 \xi +9} r_{H}^{\frac{1}{2} \left(\sqrt{8 \xi +9}+1\right)}+24 \pi  (\xi +1) r_{H} \mathrm{T}+3\right)}{48 \pi  (\xi +1)},
	\end{eqnarray}
	and, by considering $\alpha_{i}=0$ and $\Lambda=0$, we obtain
	\begin{eqnarray}
	\lim\limits_{\alpha_{i},\, \Lambda\rightarrow 0}P=\frac{4 \pi \, T\, r_{H}-1}{8 \pi  r_{H}^2}.
	\end{eqnarray}
	Finally, the heat capacity for such a deformed black hole can be obtained by
	\begin{eqnarray}
	C_{\alpha_{i}}=	T_{\alpha_{i}}\frac{	\partial S/r_{H}}{\partial T_{\alpha_{i}}/r_{H}}=\frac{2 \pi  r^{\frac{1}{2} \left(\sqrt{8 \xi +9}+1\right)} \left(-3 \sqrt{8 \xi +9}+2 \Lambda  (\xi +1) \sqrt{8 \xi +9} r^{\frac{1}{2} \left(\sqrt{8 \text{$\xi $1}+9}+1\right)}+3\right)}{\left(\sqrt{8 \xi +9}-1\right) \left(\Lambda  (\xi +1) \sqrt{8 \xi +9} r^{\frac{1}{2} \left(\sqrt{8 \xi +9}+1\right)}+3\right)}\label{43}.
	\end{eqnarray}
	Again, assuming $\alpha_{i}=\Lambda=0$, we obtain $C_{\alpha_{i}}=- 2 \pi r_{H}^{2}$.
	\begin{figure}[h!]
		\centering
		\subfigure[]{
			\begin{minipage}[t]{0.5\linewidth}
				\centering
				\includegraphics[width=3.4in,height=2.6in]{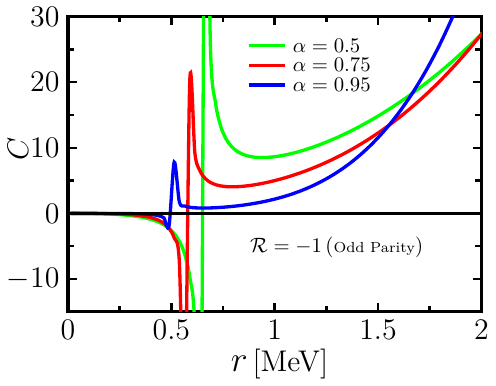}
			\end{minipage}%
		}%
		\subfigure[]{
			\begin{minipage}[t]{0.5\linewidth}
				\centering
				\includegraphics[width=3.4in,height=2.6in]{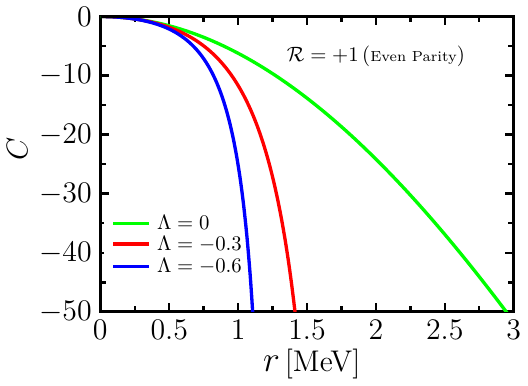}
			\end{minipage}%
		}%
		\centering
		\caption{Heat capacity vs $r$ for (a) $\mathcal{R}=-1$ (odd parity) with different values of $\alpha=0.5,0.75,0.95$, and (b) $\mathcal{R}=+1$ (even parity) with $\Lambda=0,-0.3,-0.6$.}
		\label{fig5}
	\end{figure}

In Fig. \eqref{fig5}, we plot the heat capacity as the function of radius $r$ for (a) $\mathcal{R}=-1$ (odd parity) with different values of the Dunkl parameter $\alpha=0.5,0.75,0.95$. Here, for all latter parameters, a remarkable aspect gives rise to: the appearance of a discontinuity at the critical radius, namely, $r_{1} = 0.65175205$ (green line), $r_{2}=0.572494$ (red line) and $r_{1}= 0.508152$ (blue line). In other words, we have the existence of phase transitions located in those points. On the other hand, in Fig. \eqref{fig5}(b),  we plot the heat capacity as the function of radius $r$ for $\mathcal{R}=+1$ (even parity) with $\Lambda=0,-0.3,-0.6$. The phase transition to the latter case did not occur. Moreover, some similar analyzes have recently been provided in the literature, addressing this thermodynamic function to different scenarios \cite{46,47,49,51,52,53,54,55,56}.

The deformed Schwarzschild black hole presented in this work displays thermodynamic properties that significantly deviate from the classical solution. Notably, the emergence of phase transitions in the odd parity case indicates a potential link to critical phenomena in high--energy physics. Furthermore, the modifications in thermodynamic quantities and the event horizon structure could impact observable phenomena, such as black hole shadow imaging, gravitational lensing, and \textit{Hawking} radiation.

The Dunkl parameters directly influence the thermodynamic stability of the black hole, introducing parity--specific effects that lead to qualitatively distinct behaviors. For odd parity solutions, the phase transition emerges as a result of the reflection symmetry encoded in the Dunkl operator, which modifies the thermodynamic quantities such as heat capacity and \textit{Hawking} temperature.

This divergence identifies a critical radius where the system undergoes a transition between different thermodynamic regimes. Odd parity contributions amplify the sensitivity of the thermodynamic properties to the Dunkl parameters, resulting in a more complex structure not present in the even parity case, where phase transitions are absent.
	\section{Conclusion}\label{Conclusions}
	This paper was aimed at constructing a deformed Schwarzschild black hole from the de Sitter gauge theory of gravity in presence of Dunkl terms. We determined the metric coefficients with the dependence of Dunkl parameters and parity operators. Once the spacetime coordinates were not affected by the group transformations, only the fields changed under the action of the symmetry group. Essentially, we chose a particular ansatz for the gauge fields so that the components of the strength tensor in presence of the Dunkl
	operators could be addressed.
	
	Furthermore, with such an approach, we were able to calculate the fundamental parts of theory under consideration: Riemann tensor, Ricci tensor, curvature scalar, field equations, and the integration of these equations according to Dunkl parameters and parity operators. In this sense, we presented the modifications on the
	thermodynamical properties of a spherical symmetric black hole due to the Dunkl parameter contributions for	even and odd parities. Finally, we verified a remarkable aspect highlighted from the heat capacity: the existence of a phase transition when the odd parity was invoked.
	\appendix
	
	\section{The relations of the Einstein field equations (9-14)} \label{A1}
	The non--zero components of $F_{\mu \nu}^{a}$ and $F_{\mu \nu}^{ab}$ in the presence of Dunkl operator were obtained in \cite{4} 
	\begin{equation}
	\begin{aligned}
	&F_{01}^{0}=-A'-\frac{U}{A}-\frac{A }{r}\sum_{i=1}^{3}\alpha_{i}(1-R_{i}),\quad F_{12}^{2}=r\, C'-\frac{W}{A}+\left(1+\sum_{i=1}^3 \alpha_i\left(1-R_i\right)\right) C,\\
	&F_{13}^3=\left[\left(1+\sum_{i=1}^3 \alpha_i\left(1-R_i\right)\right) C+r c^{\prime}-\frac{z}{A}\right] \sin \theta,\quad F_{02}^3=r C V,\quad F_{03}^{2}=-r V\, C\, \sin\theta.
	\end{aligned}
	\label{2.7}
	\end{equation}
	and respectively
	\begin{equation}
	\begin{aligned}
	F_{\mu \nu}^{a b}= D_\mu^{\alpha_{i}} \omega_\nu^{a b}-D_\nu^{\alpha_{i}} \omega_\mu^{a b}+\left(\omega_\mu^{a c} \omega_\nu^{d b}-\omega_\nu^{a c} \omega_\mu^{d b}\right) \eta_{c d}-4 \lambda^2\left(e_\mu^a e_\nu^b-e_\nu^a e_\mu^b\right)=R_{\mu \nu}^{a b},
	\end{aligned}
	\end{equation}
	\begin{equation}
	\begin{aligned}
	&F_{01}^{01}=-U^{\prime}-\frac{1}{r} \sum_{i=1}^3 \alpha_i\left(1-R_i\right) U-4 \lambda^2,\quad
	F_{01}^{23}=-V^{\prime}-\frac{V}{r} \sum_{i=1}^3 \alpha_i\left(1-R_i\right), \\
	&F_{23}^{13}=\left(Z+\sum_{i=1}^2 \alpha_i\left(1-R_i\right)-\alpha_3\left(1+\mathcal{R}_{3}\right)\tan^2 \theta-W\right) \cos \theta,\quad F_{02}^{13}=V W, \\
	&F_{03}^{03}=-\left(Z U+4 \lambda^2 r C A\right) \sin \theta,\quad F_{02}^{02}=-U W-4 \lambda^2 r C A, \\
	&F_{12}^{12}=W^{\prime}+\frac{W}{r} \sum_{i=1}^3 \alpha_i\left(1-R_i\right)-4 \lambda^2 r \frac{C}{A} ,\quad F_{03}^{12}=-V Z\sin \theta, \\
	&F_{13}^{13}=\left(Z^{\prime}+\frac{Z}{r} \sum_{i=1}^3 \alpha_i\left(1-R_i\right)-4 \lambda^2 r \frac{C}{A}\right) \sin \theta, \\
	&F_{23}^{23}=-\left(1-\sum_{i=1}^2 \alpha_i\left(1-R_i\right) \cot^{2}\theta+\alpha_3\left(1+R_3\right)-W Z+4 \lambda^2 r^2 C^{2}\right) \sin \theta,
	\end{aligned}
	\end{equation}
	where ${A}^{\prime },\,{U}^{\prime },\,{V}^{\prime },\,{W}^{\prime
	}$and ${Z} ^{\prime }$ denote the derivatives of first order with
	respect to the $r$ coordinate.
	
	\section{The Riemann tensor in the presence of Dunkl operator} \label{ApendixB}
	First, we calculate the Riemann tensor for our model, defining the following
	formula:
	\begin{eqnarray}
	\tilde{R}_{\mu \nu}^{\rho \sigma}=F_{\mu \nu}^{ab}e_{a}^{\rho}e_{b}
	^{\sigma}.
	\end{eqnarray}
	The corresponding non--null components in the presence of Dunkl operator are
	\begin{equation}
	\begin{aligned}
	&\tilde{R}_{01}^{01} =\left(-U^{\prime}-\frac{1}{r} \sum_{i=1}^3 \alpha_i\left(1-\mathcal{R}_{i}\right) U-4 \lambda^2\right), \\
	&\tilde{R}_{01}^{23}=-\left(V^{\prime}+\frac{V}{r} \sum_{i=1}^3 \alpha_i\left(1-\mathcal{R}_{i}\right)\right) \frac{1}{r^2 c^2 \sin ^2 \theta}, \\
	&\tilde{R}_{02}^{02}=-\frac{U W}{r A C}-4 \lambda^2,\\
	&\tilde{R}_{02}^{13} =\frac{V W A}{r C \sin \theta} ,\quad\tilde{R}_{03}^{03} =-\frac{Z V}{r C A}-4 \lambda^2 ,\quad
	\tilde{R}_{03}^{12}=-\frac{A V Z \sin \theta}{r C}, \\
	&\tilde{R}_{12}^{12} =\frac{W^{\prime} A}{r C}+\frac{W A}{r^2 C} \sum_{i=1}^3 \alpha_i\left(1-R_i\right)-4 \lambda^2, \\
	&\tilde{R}_{13}^{13} 
	=\frac{A Z^{\prime}}{r C}+\frac{A Z}{r^2 C} \sum_{i=1}^3 \alpha_i\left(1-R_i\right)-4 \lambda^2,\\
	&\tilde{R}_{23}^{13}=\left(Z+\sum_{i=1}^2 \alpha_i\left(1-R_i\right)-W\right) \frac{A \cos \theta}{r C \sin \theta}-\alpha_3\left(1+R_3\right) \frac{A \tan \theta}{r C}, \\
	&\tilde{R}_{23}^{23}=\frac{W Z-1+\alpha_3\left(1+R_3\right)}{r^2 C^2}-4 \lambda^2-\frac{1}{r^2 C^2} \sum_{i=1}^2 \alpha_i\left(1-R_i\right) \cot^{2}\theta.
	\end{aligned}
	\end{equation}
	Then, we calculate the components of the Ricci tensor defined as
	\begin{eqnarray}
	\tilde{R}_{\mu}^{\nu}=R_{\mu \rho}^{ab}e_{a}^{\nu}e_{b}^{\rho},
	\end{eqnarray}
	so that we can obtain the following non--null components:
	\begin{equation}
	\begin{aligned}
	&\tilde{R}_0^0 =-U^{\prime}-\frac{1}{r} \sum_{i=1}^3 \alpha_i\left(1-R_i\right) U-\frac{(W+Z) U}{r A C}-12 \lambda^2, \\
	&\tilde{R}_1^{1}=-U^{\prime}-\frac{1}{r} \sum_{i=1}^3 \alpha_i\left(1-R_i\right) U+\frac{\left(W^{\prime}+Z^{\prime}\right)A}{r C}+\left(\frac{W+Z}{r^2 C}\right) A \sum_{i=1}^3 \alpha_i\left(1-R_i\right)-12 \lambda^2, \\
	&\tilde{R}_2^{1} =\left(Z-W-\sum_{i=1}^2 \alpha_i\left(1-R_i\right)\right) \frac{A \cos \theta}{r C \sin \theta}-M_3\left(1+R_3\right) \frac{A \tan \theta}{r C},
	\end{aligned}
	\end{equation}
	\begin{equation}
	\begin{aligned}
	&\tilde{R}_2^2=\frac{W Z-1+\alpha_3\left(1+R_3\right)}{r^2 C^2}-12 \lambda^2-\frac{1}{r^2 C^2} \sum_{i=1}^2 \alpha_i\left(1-R_i\right) \cot\theta^2+\frac{W^{\prime} A}{r C}+\\&\frac{W A}{r^2 C} \sum_{i=1}^3 M_i\left(1-R_i\right)-\frac{U W}{r C A}, \\
	&\tilde{R}_3^3
	=\frac{-Z U}{r C A}+\frac{A Z^{\prime}}{r C}-\frac{1-Z W-\alpha_3\left(1+R_3\right)+\sum_{i=1}^2 \alpha_i\left(1-R_i\right) \cot\theta^2}{r^2 C^2}+\\&\frac{A Z}{r^2 C} \sum_{i=1}^3 \alpha_i\left(1-R_i\right)-12 \lambda^2.
	\end{aligned}
	\end{equation}
	In order to write the Einstein equations, we calculate the Ricci scalar $\tilde{R}=R_{\mu}^{\mu}$
	\begin{equation}
	\begin{aligned}
	\tilde{R}=&-2\left(U^{\prime}+\frac{1}{r} \sum_{i=1}^3 \alpha_i\left(1-R_i\right) U+\frac{(W+Z)}{r C A} U- \frac{\left(W^{\prime}+Z^{\prime}\right)}{r C} A+24 \lambda^2\right. \\
	&\left.-\frac{(W+Z) A}{r^2 C} \sum_{i=1}^3 \alpha_i\left(1-R_i\right)+\frac{\left(1-W Z-\alpha_3\left(1+R_3\right)+\sum_{i=1}^2 \alpha_i\left(1-R_i\right) \cot\theta^2\right)}{r^2 C^2}\right).
	\end{aligned}
	\end{equation}
	The Einstein equations for the vacuum can be written as
	\begin{eqnarray}
	\tilde{R}_{\mu}^{\nu}-\frac{1}{2}\delta_{\mu}^{\nu}\tilde{R}=0.
	\end{eqnarray}
	For the above model, we have
	\begin{eqnarray}
	\text{(i)}\,\,&&\tilde{R}_{0}^{0}-\frac{1}{2}\delta_{0}^{0}\tilde{R}=0,	\\
	&&-\frac{\left(W^{\prime}+Z^{\prime}\right) A}{r C}+\frac{\left(1-W Z-\alpha_3\left(1+R_3\right)+\sum_{i=1}^2 \alpha_i\left(1-R_i\right) \cot\theta^2\right)}{r^2 C^2}\nonumber\\&&+\frac{(W+Z) A}{r^2 C} \sum_{i=1}^3 \alpha_i\left(1-R_i\right)+12 \lambda^2=0.\nonumber
	\end{eqnarray}
	\begin{eqnarray}
	\text{(ii)}\,\,&&\tilde{R}_{1}^{1}-\frac{1}{2}\delta_{1}^{1}\tilde{R}=0,	\\&&
	\frac{(W+Z) U}{r C A}+\frac{1-W Z-\alpha_3\left(1+R_3\right)+\sum_{i=1}^2 \alpha_i\left(1-R_i\right)\cot \theta^{2}}{r^2 C^2}+12 \lambda^2=0.\nonumber
	\end{eqnarray}
	\begin{eqnarray}
	\text{(iii)}\,\,	&&\tilde{R}_{2}^{2}-\frac{1}{2}\delta_{2}^{2}\tilde{R}=0,	\\&&
	U^{\prime}+\frac{1}{r} \sum_{i=1}^3 \alpha_i\left(1-R_i\right) U+\frac{Z U}{r C A}-\frac{Z^{\prime} A}{r C}-\frac{Z A}{r^2 C} \sum_{i=1}^3 \alpha_i\left(1-R_i\right)+12 \lambda^2=0.\nonumber\end{eqnarray}
	\begin{eqnarray}
	\text{(iv)}\,\,	&&\tilde{R}_{3}^{3}-\frac{1}{2}\delta_{3}^{3}\tilde{R}=0,	\\&&
	U^{\prime}+\frac{1}{r} \sum_{i=1}^3 \alpha_i\left(1-R_i\right) U+\frac{W U}{r C A}-\frac{W^{\prime} A}{r C}-\frac{W A}{r^2 C} \sum_{i=1}^3 \alpha_i\left(1-R_i\right)+12 \lambda^2=0.\nonumber\end{eqnarray}
	\begin{eqnarray}
	\text{(v)}\,\,&&\tilde{R}_{2}^{1}-\frac{1}{2}\delta_{2}^{1}\tilde{R}=0,	\\&&(W-Z)A+\sum_{i=1}^3 \alpha_i\left(1-R_i\right) \frac{A \cos \theta}{r C \sin \theta}+\alpha_3\left(1+R_3\right) \frac{A \tan \theta}{r C}=0.\nonumber
	\end{eqnarray}
    
	\section*{Acknowledgements}

    A. A. Araújo Filho is supported by Conselho Nacional de Desenvolvimento Cient\'{\i}fico e Tecnol\'{o}gico (CNPq) and Fundação de Apoio à Pesquisa do Estado da Paraíba (FAPESQ) -- [200486/2022-5] and [150891/2023-7].

\section{Data Availability Statement}

Data Availability Statement: No Data associated in the manuscript.


\end{document}